\documentclass[iicol,pdflatex,sn-mathphys-num]{sn-jnl}%
\usepackage{geometry}
\usepackage{graphicx}
\usepackage{multirow}
\usepackage{amsmath,amssymb,amsfonts}
\usepackage{amsthm}
\usepackage{mathrsfs}
\usepackage[title]{appendix}
\usepackage{xcolor}
\usepackage{textcomp}
\usepackage{manyfoot}
\usepackage{booktabs}
\usepackage{siunitx}
\usepackage{tikz}
\usetikzlibrary{arrows.meta, positioning, fit, calc}

\begin{document}

\title[mode=title]{Anomaly Detection in Bunch Oscillations for Early Warning of Sudden Beam Loss at SuperKEKB}

\author[1]{\fnm{H.} \sur{Haigh}}\email{huw.haigh@oeaw.ac.at}
\author[2]{\fnm{P.} \sur{Horak}}\email{phorak@uvic.ca}
\author[1]{\fnm{G.} \sur{Inguglia}}\email{Gianluca.Inguglia@oeaw.ac.at}
\author[1]{\fnm{C.} \sur{Krause}}\email{Claudius.Krause@oeaw.ac.at}
\author[3,5]{\fnm{G.} \sur{Mitsuka}}\email{gaku.mitsuka@kek.jp}
\author[3,5]{\fnm{H.} \sur{Nakayama}}\email{hiroyuki.nakayama@kek.jp}
\author[3,4]{\fnm{R.} \sur{Nomaru}}\email{nomaru@post.kek.jp}
\author[3,5]{\fnm{M.} \sur{Tobiyama}}\email{makoto.tobiyama@kek.jp}
\affil[1]{\orgname{Marietta Blau Institute, Austrian Academy of Sciences}, \orgaddress{\city{Vienna}, \country{Austria}}}
\affil[2]{\orgname{University of Victoria}, \orgaddress{\city{Victoria}, \state{British Columbia}, \country{Canada}}}
\affil[3]{\orgname{High Energy Accelerator Research Organization (KEK)}, \orgaddress{\city{Tsukuba}, \country{Japan}}}
\affil[4]{\orgdiv{Department of Physics}, \orgname{The University of Tokyo}, \orgaddress{\city{Tokyo}, \country{Japan}}}
\affil[5]{\orgname{The Graduate University for Advanced Studies (SOKENDAI)}, \orgaddress{\city{Hayama}, \state{Kanagawa}, \country{Japan}}}
\abstract{\unboldmath Sudden beam loss (SBL) events at the SuperKEKB collider result in the loss of a large fraction of the stored beam within just a few revolutions, on a timescale comparable to or shorter than the $20$--$30\,\mu\mathrm{s}$ latency of the existing abort system. Such events have damaged collimators and threaten Belle~II subdetectors. Bunch Oscillation Recorders (BORs) capture transverse bunch positions at each revolution, which often show sizable oscillations that precede significant beam current loss. We test whether these precursors can be detected earlier than a charge-based reference that mimics the existing abort sensors. We apply several oscillation-based anomaly detection algorithms to 208 SBL events recorded in the low-energy ring during the 2024 run period, ranging from simple rolling statistics to machine learning models trained to forecast bunch oscillation. On 173 events not used for training or calibration, the best-balanced model (DeepLSTM) triggers before the charge-based reference in 86--90\% of events across five training seeds, with a mean lead time of 25--30\,$\mu$s. The only trigger classified as a false alarm fires on an oscillation that begins about 60 turns before the abort and persists until the beam is lost, making it more likely an early precursor than a spurious trigger. These results establish bunch oscillations as a viable early abort trigger and motivate a shadow-mode deployment to measure the false-alarm rate in continuous operation.}


\maketitle

\section{Introduction}\label{sec:intro}
Sudden beam loss (SBL) events are a major operational risk at the SuperKEKB electron-positron collider.
During such events, a large fraction of the stored beam is lost within a period of tens of microseconds, resulting in large doses of radiation being delivered to sensitive instrumentation. Previous SBL events have damaged collimator heads and about 2\% of the gates of the Belle~II pixel detector~\cite{terui2026sbl}. Since then, physics data has been recorded with the pixel detector switched off~\cite{terui2026sbl}. The increasing SBL frequency at high beam currents~\cite{ikeda2025ipac} also limits the integrated luminosity that can be delivered to Belle~II, whose physics program is built around a target data set of $50\,\mathrm{ab}^{-1}$~\cite{Kou:2018nap}.

The existing abort system mainly uses signals that come after beam loss has started. Once a trigger is issued, a further $20$--$30\,\mu\mathrm{s}$ is required to dump the beam~\cite{yoshihara2024}. Because an SBL can develop on a comparable or shorter timescale, an earlier indicator is needed to prevent or reduce damage.

Bunch Oscillation Recorders (BORs) measure transverse particle bunch positions in each revolution. Previous observations show that sizable transverse oscillations often develop before significant charge loss during SBL events~\cite{nomaru2024,nomaru2025}. These oscillations may therefore provide an early warning, allowing an abort to be initiated faster than conventional loss-based sensors. Similar machine-learning approaches have been used for early detection of accelerator faults~\cite{wielgosz2017lstm,blokland2022errant,rajput2024errant,tennant2023smart,suetsugu2024pressure}, while this work applies the approach to BOR-based SBL precursors at SuperKEKB.

In this work, we test whether these precursors can be detected early and reliably in BOR data. We compare several anomaly detection methods, ranging from rolling statistics to multivariate forecasting models, on 208 SBL events recorded during the 2024 run. Each method is evaluated by its time advantage over a charge-based reference and by its number of false alarms. This offline study aims to identify the most promising models for online implementation, where the false-alarm rate in continuous operation can be measured (Section~\ref{sec:fa_rate}).

Section~\ref{sec:superkekb_accel} introduces the SuperKEKB abort system and BOR instrumentation, Section~\ref{sec:sample} describes the dataset, and Sections~\ref{sec:methods}--\ref{sec:results} present the detection methods, optimisation procedure, and results. Deployment considerations and conclusions are discussed in Sections~\ref{sec:caveats} and~\ref{sec:summary}.
\section{SuperKEKB and Sudden Beam Losses}\label{sec:superkekb_accel}

\begin{figure}[h]
  \centering
  \includegraphics[width=\linewidth]{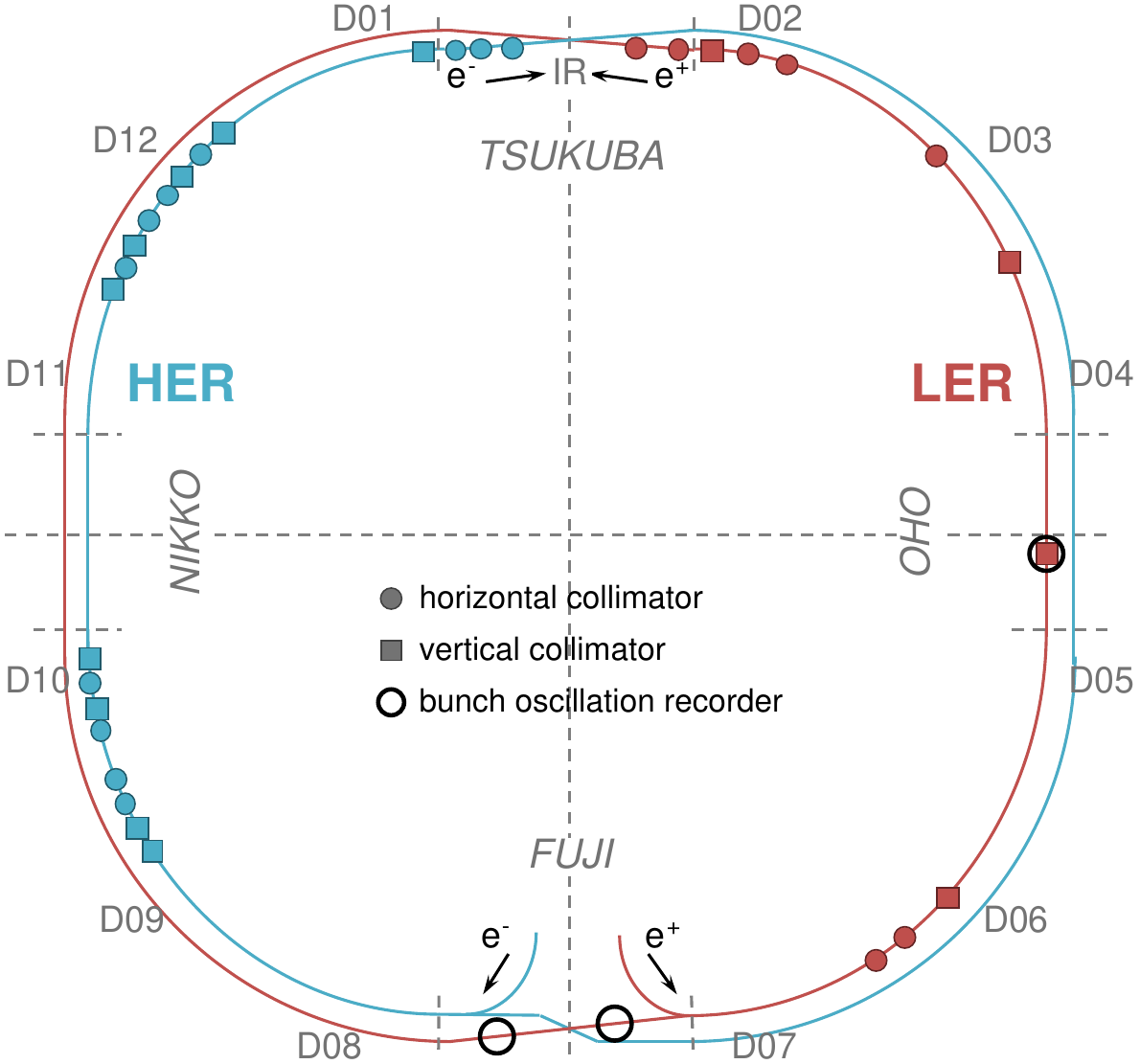}
  \caption{Layout of the SuperKEKB main ring, showing the HER and LER, the interaction point (IP), the collimator locations, and the accelerator sections. Markers indicate the BOR stations in the Fuji and D05 sections of the LER.}
  \label{fig:superkekb_layout}
\end{figure}

The SuperKEKB collider at KEK in Tsukuba, Japan, is an energy-asymmetric electron--positron collider operating at the $\Upsilon(4S)$ resonance of $10.58\,\mathrm{GeV}$, with collisions optimized for the production of $B$ mesons, $\tau^+ \tau^-$, and $c\bar{c}$ events. It employs the nano-beam collision scheme~\cite{Akai:2018mbz}, which achieves high luminosity by tightly squeezing the vertical beam size at the interaction point (characterized by $\beta^*_y$, currently operating at $1\,\mathrm{mm}$), to reach a target luminosity of $6\times10^{35}\,\mathrm{cm^{-2}\,s^{-1}}$~\cite{belle2_backgrounds}. It consists of two storage rings of $3\,\mathrm{km}$ circumference: a high-energy ring (HER) storing $7\,\mathrm{GeV}$ electrons and a low-energy ring (LER) storing $4\,\mathrm{GeV}$ positrons, as shown in Fig.~\ref{fig:superkekb_layout}. As of June~2026, the collider has reached a peak instantaneous luminosity of $5.3\times10^{34}\,\mathrm{cm^{-2}\,s^{-1}}$~\cite{skekb_dailysnap}, the highest ever achieved by a particle collider.

The accelerating radio frequency (RF) cavities operate at $509\,\mathrm{MHz}$, providing 5120 equally spaced bucket positions around each ring, of which approximately 2300 are filled with bunches during typical operation. The bunches are arranged in two trains separated by abort gaps of approximately $200\,\mathrm{ns}$. A single revolution (\textit{turn}) takes approximately $10\,\mu\mathrm{s}$. Continuous top-up injection at rates up to $25\,\mathrm{Hz}$ per beam maintains beam currents in excess of $1\,\mathrm{A}$ in both rings~\cite{Akai:2018mbz}.

The Belle~II experiment~\cite{Abe:2010gxa} is located at the single interaction point (IP) where the beams collide at a crossing angle of $83\,\mathrm{mrad}$. The innermost pixel detector (PXD), silicon vertex detector (VXD), and several other components are sensitive to radiation damage, making beam loss control a critical operational requirement. A system of movable collimators~\cite{ishibashi2020collimator} blocks stray particles from reaching the detector. Their apertures can be as small as $1\,\mathrm{mm}$ and are controlled with a precision of $50\,\mu\mathrm{m}$.

\subsection{Sudden Beam Losses}\label{sec:sbl}

Beam losses occur regularly during collider operation and can be initiated by several mechanisms including Touschek scattering, beam-gas interactions, and beam-beam effects~\cite{belle2_backgrounds}. These losses develop gradually, thereby allowing abort sensors to detect them early enough and the beam to be safely dumped before significant damage occurs. In contrast, during SBL events, a large fraction of the stored beam is lost within a few turns, or tens of microseconds. These timescales are too short for safe abort, and the resulting radiation burst can deposit several thousand mrad on sensitive detector components and collimators in a single event. SBLs have caused damage to collimator heads and especially threaten the pixel detector~\cite{ikeda2025ipac,yoshihara2024}. Beyond direct detector damage, each abort requires refilling the beam from zero, and the increasing frequency of SBLs at higher currents constrains the accumulation of integrated luminosity. The most severe events can trigger quenches of the superconducting final-focus magnets (QCS), resulting in significant downtime for magnet recovery.

The frequency of SBLs increased during 2024 operations as beam currents surpassed those typical of its predecessor, KEKB~\cite{Abe:2013ftp}, reaching more than five events per day at the highest currents~\cite{terui2026sbl}. They occur predominantly in the LER: during the 2024 run, SBL-type aborts accounted for 16\% of all LER aborts compared to 1.4\% in the HER~\cite{ikeda2025ipac}. They have been observed under single-beam conditions (without collisions), at multiple optics configurations ($\beta^*_y = 1$, 3~mm, and detuned), and with both 2- and 3-bucket-spacing filling patterns, suggesting that the cause is not specific to a single operational mode. The leading hypothesis is an interaction between the beam and dust. SBLs were reproduced artificially by knocking on beam pipes and by discharging clearing electrodes, and black stains found at flange joints were identified as degraded vacuum sealant~\cite{terui2026sbl}. Simulations show that dust in the beam can cause a rapid growth of the vertical beam size that leads to beam loss~\cite{ohmi2026dust}, although a complete understanding of SBL has not yet been reached~\cite{terui2026sbl}.

Vacuum sealant was removed from flange joints throughout the LER ahead of the 2025 and 2026 run periods~\cite{terui2026sbl}. After operations resumed in 2025, the rate of SBLs decreased. However, SBLs have continued to occur, with 102 events recorded over 155 operating days in the 2025 and 2026 run periods, 37 of which led to QCS quenches. SBLs therefore remain an operational concern, and mitigating their impact is important regardless of their underlying cause. This paper focuses on mitigation rather than on identifying the underlying cause of SBLs.

\subsection{Abort systems}\label{sec:abort}

When an abort sensor detects an anomaly, kicker magnets deflect the circulating beam toward a dedicated beam dump~\cite{mimashi2014abort}. Trigger signals are timed to the abort gaps between the two bunch trains, so the beam is dumped safely. The total latency between anomaly detection and beam dump is typically 20--30\,$\mu\mathrm{s}$, depending on the sensor location and the distance between the lost bunch and the nearest abort gap~\cite{yoshihara2024}.

Abort sensors are distributed around both rings. Loss monitors, based on PIN photodiodes and optical fibres, are installed around the ring and predominantly near collimators, where losses are most likely due to the narrow aperture~\cite{ikeda2014blm,yoshihara2024}. Single-crystal CVD diamond sensors are mounted on the beam pipe near the interaction point, on SVD support cones, and on the QCS bellows. They provide high-speed, localised coverage of the IP region and contribute to abort triggering by acting as solid-state ionization chambers~\cite{bacher2021diamond}. The CLAWS (sCintillation Light And Waveform Sensors) system comprises 32 plastic-scintillator/SiPM modules arranged symmetrically around the IP at distances of $1$--$4\,\mathrm{m}$ and at four azimuthal angles~\cite{yoshihara2024}. The CLAWS system was originally used to monitor injection-induced backgrounds, but because of its rapid response relative to other detectors in some cases, it was repurposed as a primary abort sensor after 2020.

\subsection{Bunch Oscillation Recorders}
\begin{figure}[h]
    \centering
    \includegraphics[width=\linewidth]{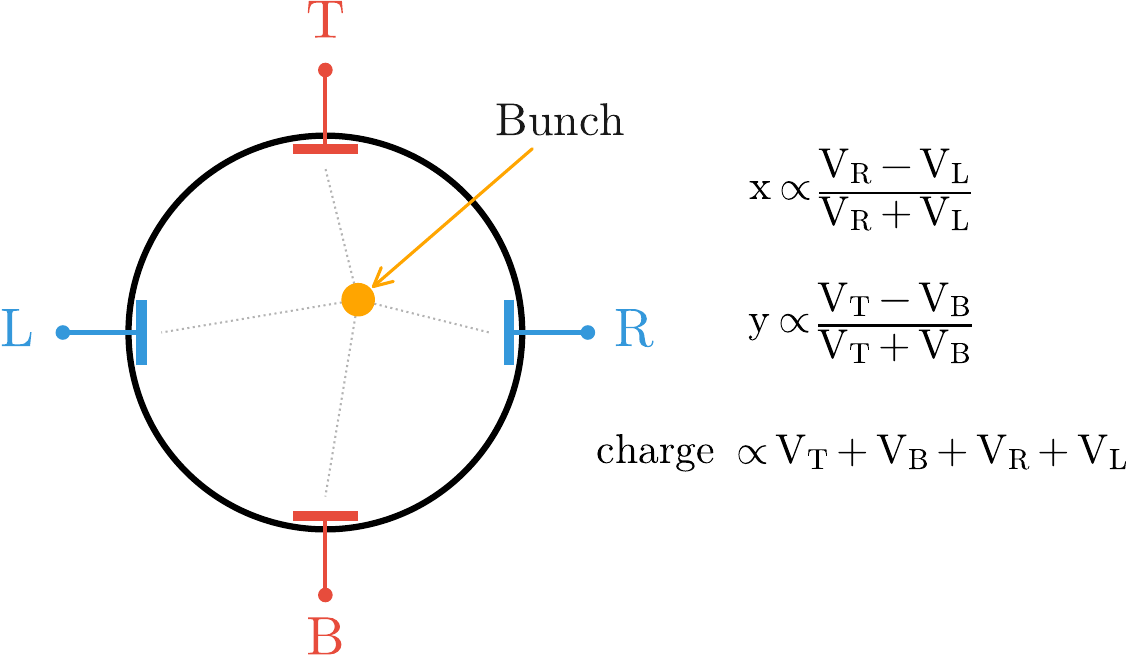}
    \caption{Schematic of a beam position monitor. Four button electrodes surround the beam pipe. The difference of the signals on opposing electrodes is proportional to the transverse displacement multiplied by the bunch charge, and their sum is proportional to the bunch charge. The BORs record these signals turn by turn and the normalization by the bunch charge is applied offline.}
    \label{fig:bpm-schematic}
\end{figure}
\begin{figure}[h]
  \centering
  \includegraphics[width=\linewidth]{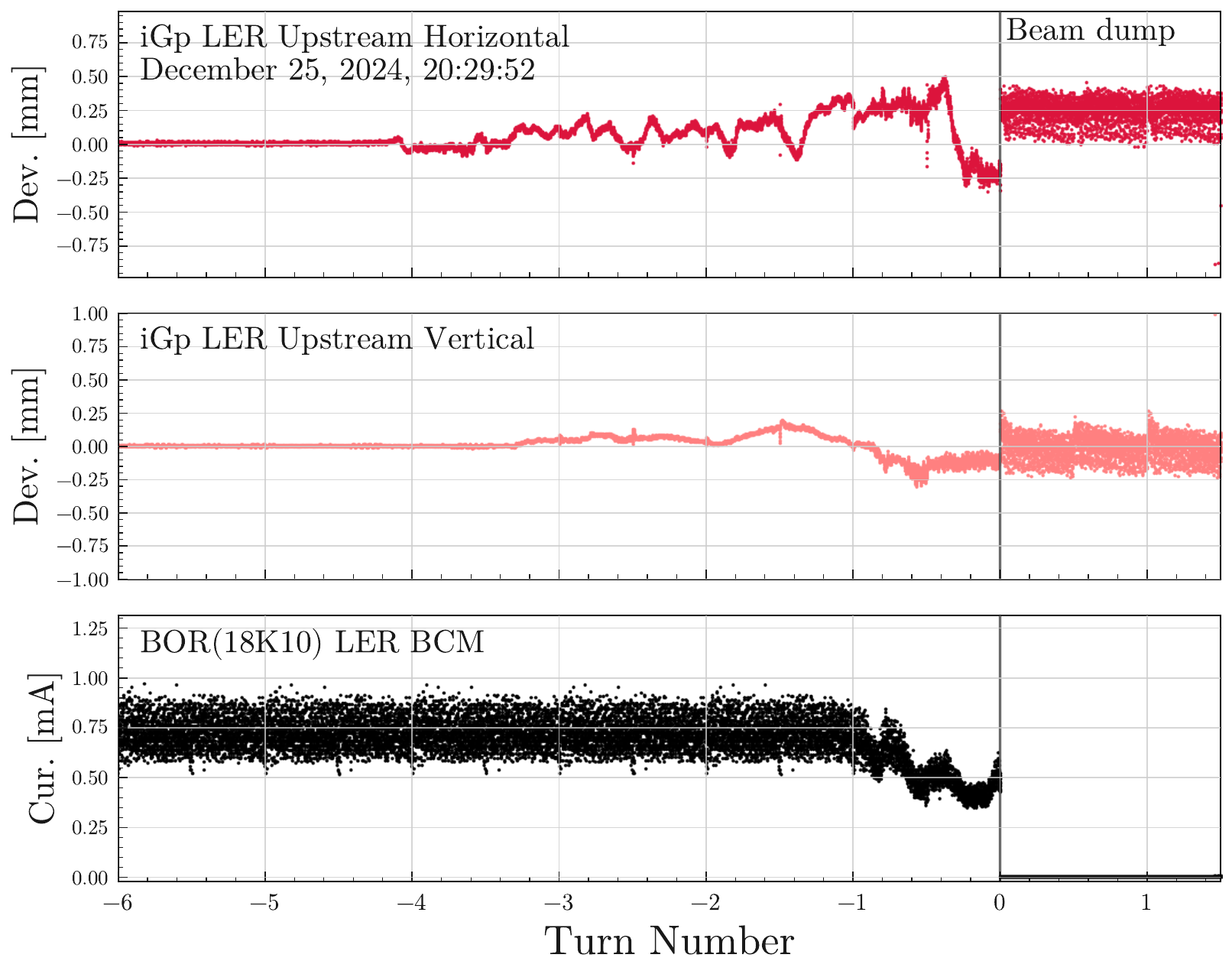}
  \caption{Example recording of an SBL event, from the LER on 25 December 2024. The upper two panels show the turn-by-turn horizontal and vertical bunch position recorded by the upstream iGp12-based BOR, the lower panel the bunch charge from the BCM channel. The oscillation amplitude grows while the charge is still at its nominal level.}
  \label{fig:bor_example}
\end{figure}

Since SBLs can develop faster than the 20--30\,$\mu\mathrm{s}$ abort latency~\cite{ikeda2025ipac,yoshihara2024}, characterizing the beam dynamics in the moments before a loss is highly valuable. A beam position monitor (BPM) measures the transverse displacement of a passing bunch by comparing the induced charges on four button electrodes surrounding the beam pipe, from which the center position is inferred (Fig.~\ref{fig:bpm-schematic}). Summing the charges across all four electrodes provides a measure of the bunch-by-bunch charge, a functionality referred to as a bunch current monitor (BCM). A Bunch Oscillation Recorder (BOR) continuously records this information turn-by-turn in a buffer, which is saved upon receipt of the abort kicker signal. Early measurements showed that SBL events are often preceded by sizable transverse oscillations before significant charge is lost, making BOR data the earliest available indicator of an impending SBL~\cite{nomaru2024,nomaru2025}. While Ref.~\cite{nomaru2024} focuses on the precursor phenomenon and the recording hardware, this paper focuses on automated detection and triggering. Figure~\ref{fig:bor_example} shows a representative recording, in which the transverse position begins to oscillate before the charge decreases.
Three BOR types are in operation at SuperKEKB, two of which provide the data used in this study. The conventional BOR (ADC)~\cite{tobiyama2012bor}, based on the 18K10 VME board by Digitex, is installed in the Fuji section of each ring (marked in Fig.~\ref{fig:superkekb_layout}). The iGp12-based BOR uses the iGp12 signal processor (Dimtel~\cite{teytelman2006gproto}) of the bunch-by-bunch feedback system~\cite{tobiyama2016} as a readout, providing two additional BPM stations (upstream and downstream) in the Fuji section. The newest BOR, based on Radio Frequency System-on-Chip (RFSoC), is a fully independent instrument combining a new BPM pickup with an AMD/Xilinx RFSoC for digitization and processing. It achieves better position resolution than the ADC while being compact enough to deploy at multiple locations~\cite{nomaru2024}. During the 2024 run, two RFSoC units were installed in the LER, one in the Fuji area and one in the D05 section~\cite{nomaru2025}. RFSoC units for the HER are planned for 2027.

\section{Datasets}\label{sec:sample}

We focus this study on the LER, where the larger event sample (208 events, versus 38 in the HER) gives a more reliable basis for evaluating the method. The dataset consists of SBL events recorded during the 2024 run period, split into two sub-periods referred to as 2024ab and 2024c. Of the 220 LER events in the dataset, 12 are set aside due to processing issues with the raw data, or because they coincide with injection, during which some bunches are significantly displaced. This leaves 208 events used in the analysis.

For each event, the BOR system saves a continuous turn-by-turn buffer that was being accumulated at the time the abort was triggered. The buffer is written to disk after receiving the kicker signal, recording the beam dynamics in the turns leading up to the loss. Two BOR types contribute data to this study. The iGp12-based BOR records four position channels, upstream and downstream BPM stations in the Fuji section, each in the horizontal and vertical plane, with a buffer depth of approximately 100 turns. The ADC records the downstream position channels at a coarser time resolution, enabling a larger buffer of approximately 4000 turns. The 18K10 ADC board also records the sum signal, providing a bunch current monitor channel with the turn-by-turn bunch charge, which enables direct measurement of beam loss dynamics. The resulting dataset therefore contains six beam-position channels, covering the horizontal and vertical planes of the ADC and the upstream and downstream iGp12, together with one current channel from the ADC. The position resolution is about $30\,\mu$m for the ADC and about $2\,\mu$m for the BPM read out by the iGp12~\cite{nomaru2024}.

\begin{figure}[h]
  \centering
  \includegraphics[width=\linewidth]{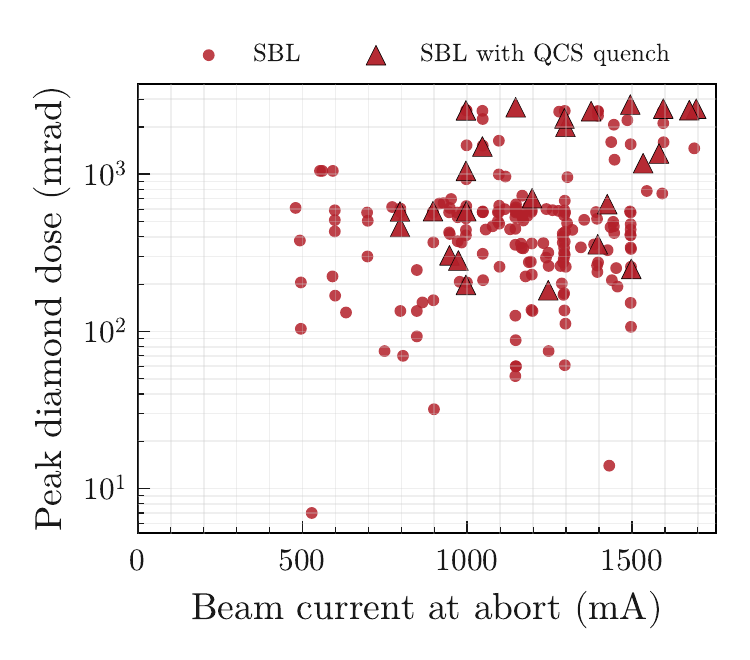}
    \caption{Peak diamond-sensor dose at the IP versus beam current at the time of the abort, for the SBL events in the LER. Triangles highlight events that led to a QCS quench. Beam currents span approximately 480 to 1700\,mA.}
  \label{fig:dataset_currents}
\end{figure}

The beam currents at the time of each abort are shown in Fig.~\ref{fig:dataset_currents}. Events span a wide range from approximately 480 to 1700\,mA, with a median of 1166\,mA, reflecting the broad range of operating conditions over which SBLs occur in this ring. The dataset was collected under standard top-up operation with continuous injection.
\section{Methods and Implementation}\label{sec:methods}

The goal of this study is to determine whether anomalous beam dynamics recorded by the BOR system can be detected early enough to provide useful warning ahead of the abort trigger, and whether such precursor signatures can be consistently detected across events. We apply three types of anomaly detection methods to the turn-by-turn BOR data: a sliding-window variance detector, a Kalman filter with a constant-velocity model~\cite{kalman1960}, and a neural-network forecaster. We refer to the first two, which score each channel independently before pooling, as \textit{classical detectors}, and to the LSTM-based forecasters, which process channels jointly through one network, as \textit{multivariate models}. Each method is described below, followed by the common evaluation procedure.

All methods take the six position channels as input. The BCM channel is not used as an input to these models but serves as the reference anchor for lead time evaluation. A rolling-variance detector applied to the BCM provides a proxy for the currently in place charge-loss-based abort sensors distributed around the ring, and all lead times in this study are measured relative to this BCM reference detection. The reference is not a perfect proxy, as loss monitors around the ring respond to charge deposition locally as it occurs, while the BCM measures a charge reduction per turn integrated over all bunches. The BCM may therefore either overestimate or underestimate the detection capability of the distributed loss monitor network by up to one turn, depending on where along the ring the charge is deposited relative to the monitor locations. This, however, does not affect the relative ordering of the detection times between the different methods.

The sliding-window and Kalman models have no learned parameters. A set of events drawn from the dataset is only used to characterize the typical anomaly score in quiet regions and set the detection threshold. For the LSTM, a separate set of events is used as training data from which the model learns to forecast bunch positions. Each training event is divided into sequences consisting of a fixed-length window of consecutive bunches, with the bunch immediately following the window used as the prediction target. To avoid exposing the model to precursor dynamics during training, only sequences ending more than 30 turns before the abort are retained. Of these, a random subset is used to keep the training computationally manageable.

The detection threshold is then calibrated in the same way for all models. The anomaly scores from each calibration event during stable beam conditions are pooled, and the threshold is set as $\tau = \mu_\text{cal} + k\,\sigma_\text{cal}$, where $\mu_\text{cal}$ and $\sigma_\text{cal}$ are the mean and standard deviation of the pooled scores, and $k$ is a dimensionless sensitivity parameter. Rather than fixing $k$ to a single value, we sweep it across a grid to characterize the trade-off between the false alarm rate and the mean lead time. Results are presented as a function of the number of false alarms per model. All anomaly scores are smoothed before threshold comparison with an equally-weighted one-sided rolling mean $\langle \cdot \rangle_W$ over the most recent $W$ samples, using only past data at each step. The smoothing window and the required number of consecutive samples above threshold before a trigger is declared are model-specific and stated in the respective subsections.

The lead time for a given event is defined as $\Delta t = t_\text{ref} - t_\text{model}$, where $t_\text{ref}$ is the BCM reference detection time and $t_\text{model}$ is the first time the model's smoothed score exceeds $\tau$, expressed in microseconds ($1\,\text{turn} = 10\,\mu\text{s}$). Events in which the model does not trigger before the reference are scored with a lead time of zero. In practice, the charge-based monitors would still trigger an abort in such cases, and the oscillation-based models are only beneficial when they fire first.

\subsection{Sliding-window variance detector}\label{sec:baseline}

For each position channel, a rolling variance is computed over a window of 500 samples. Each sample corresponds to one filled bunch, so the window spans roughly 0.2 turns for a typical fill pattern. The raw variance is normalized by the median variance during stable beam conditions, estimated from the first 50\,000 samples of the recording, giving a score that is approximately unity during nominal operation and rises when oscillation amplitude or character changes (fixed parameters for all models are collected in Table~\ref{tab:params}). The per-channel scores are combined into a single scalar anomaly score by pooling across channels. Three pooling functions are evaluated: the maximum, the mean, and the root-mean-square across the six channels. The choice of pooling function is optimized jointly with the detection threshold. A rolling mean of 500 bunches is applied before threshold comparison.

An alternative score is also evaluated, based on a rolling amplitude ratio. For each channel, the score is computed as
\begin{equation}
\label{eq:rolling_norm}
    s_i(t) = \frac{\langle |x_i| \rangle_{N_b}(t)}{\langle |x_i| \rangle_{10\,N_b}(t)},
\end{equation}
where $N_b$ is the number of filled bunches per turn. During stable operation this ratio is approximately unity; sustained oscillation growth causes the short-window mean to exceed the long-window baseline and the ratio rises. This variant does not require a separate normalization step, as the baseline is tracked continuously by the denominator. The same pooling and smoothing procedure is applied, with the pooling function also optimized jointly with the detection threshold, and a minimum run of ten consecutive samples above threshold required before a trigger is declared.

\subsection{Kalman filter}\label{sec:kalman}
\begin{figure*}[t]
    \centering
    \includegraphics[width=\linewidth]{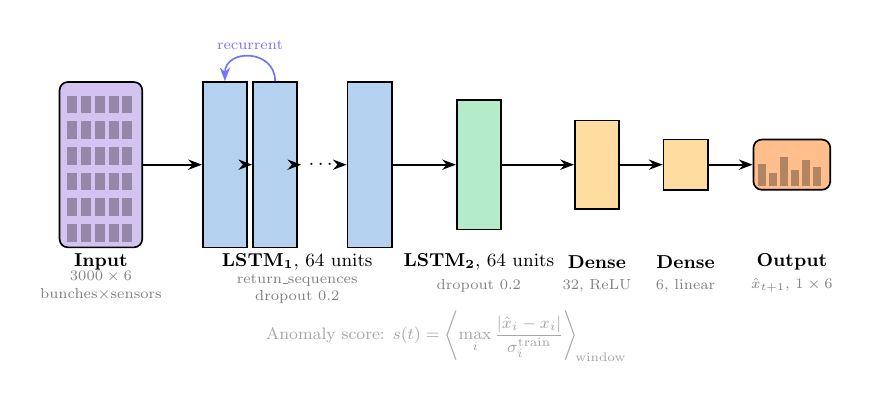}
    \caption{Architecture of the DeepLSTM forecasting model.}
    \label{fig:lstm_arch}
\end{figure*}
The Kalman filter~\cite{kalman1960} models each position channel as a system evolving under a constant-velocity prior. The state vector $\mathbf{x}_k = (p_k,\, v_k)^\top$ comprises the bunch position and its turn-to-turn rate of change, with transition and observation models
\begin{equation}
    \mathbf{x}_{k} = \begin{pmatrix} 1 & 1 \\ 0 & 1 \end{pmatrix} \mathbf{x}_{k-1} + \mathbf{w}_k, \qquad y_k = \begin{pmatrix} 1 & 0 \end{pmatrix} \mathbf{x}_{k} + \epsilon_k,
\end{equation}
where $\mathbf{w}_k \sim \mathcal{N}(0,\,Q\,\mathbf{I})$ and $\epsilon_k \sim \mathcal{N}(0,\,R)$ are process and measurement noise. Under nominal conditions the velocity state is small and slowly varying. When the beam enters a rapid oscillation the velocity estimate grows, making the magnitude of the estimated velocity state a natural anomaly score. A 10-bunch sliding-window mean of the velocity magnitude is computed for each channel, and the six channel scores are combined by pooling across channels. Similar to the sliding-window detector, three pooling functions are evaluated (maximum, mean, and root-mean-square) and the choice is optimized jointly with the detection threshold. A rolling mean of 500 bunches is applied to the pooled score before threshold comparison. A minimum run of 50 consecutive samples above threshold is required before a trigger is declared, since the velocity estimate reacts sharply to isolated noisy bunches and 50 samples ($\approx 0.2\,\mu$s) cost negligible lead time.

The process noise $Q = 10^{-5}$ and measurement noise $R = 10^{-4}$ were determined by grid search on the stable, oscillation-free portion of each calibration event to optimize tracking accuracy under nominal beam conditions by minimizing the normalized innovation squared, a filter-consistency performance metric~\cite{mehra1971innovations}.

\subsection{DeepLSTM forecasting model}\label{sec:lstm}

A DeepLSTM~\cite{hochreiter1997lstm, graves2013deep} model is trained on the LER data, taking all six position channels as a multivariate input. At each step the model receives a window of 3000 consecutive bunches (approximately 1.3 turns) and predicts the position of all six channels at the following bunch. The sequence length is chosen to exceed one full turn so that the model sees each individual bunch more than once within a single input window. The data is a flat sequence of bunches ordered by turn, so the $(N+1)^\text{th}$ entry corresponds to the same bunch as the first, the $(N+2)^\text{th}$ to the same as the second, and so on, for $N$ filled bunches per turn. Input and output values are standard-scaled to zero mean and unit variance using statistics computed from the training data, and the predictions are rescaled back to physical units before the anomaly score is computed.

The architecture, illustrated in Fig.~\ref{fig:lstm_arch}, consists of two stacked LSTM layers~\cite{hochreiter1997lstm} each with 64 units, followed by a fully connected layer with 32 units (ReLU activation) and a linear output layer producing six values. Dropout~\cite{srivastava2014dropout} with rate 0.2 is applied after each LSTM layer during training. The architecture and training hyperparameters were selected by a grid search over the training events, ranking configurations by the one-step-ahead forecasting error (mean absolute error) on a held-out validation subset. The model is trained with the Adam optimizer~\cite{Kingma2014Adam} at a learning rate of $10^{-4}$, using minibatches of 256 sequences, for up to 50 epochs with early stopping~\cite{prechelt1998earlystopping} when the validation loss does not improve for five consecutive epochs.

The per-channel anomaly score is the absolute prediction residual normalized by the per-channel standard deviation of the training data $\sigma_i^\text{train}$, clipped at 30 to limit the pull of rare extreme residuals. The combined anomaly score is obtained by pooling across channels and applying a rolling mean,
\begin{equation}
    s(t) = \left\langle \mathrm{pool}_i \frac{|\hat{x}_i - x_i|}{\sigma_i^\text{train}} \right\rangle_{500},
\end{equation}
where the subscript gives the smoothing window in bunches. The pooling function is optimized jointly with the detection threshold. A single sample above threshold is sufficient to declare a trigger.

\subsubsection{Hybrid Conv1D-LSTM model}\label{sec:cae}
\begin{figure}[h]
  \centering
  \includegraphics[width=\linewidth]{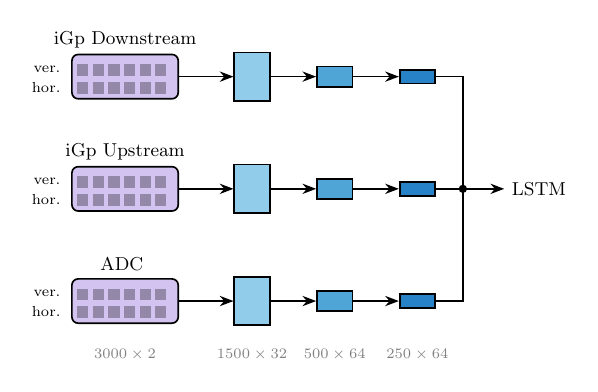}
  \caption{Architecture of the 3 stacked Conv1D layers added in the hybrid model, with data dimensions noted below each step.}
  \label{fig:cnn_architecture}
\end{figure}
Real-time deployment requires the anomaly score to be evaluated at the incoming BOR data rate of approximately 200 MHz. While effective offline, DeepLSTM processes the full 3000-sample window through two stacked recurrent layers, which is unlikely to reach this rate in its current form. Reducing either the network size or the input resolution could speed up inference, at the risk of lower detection performance. We therefore replace the first LSTM layer with one-dimensional convolutional layers~\cite{lecun1998gradient} that shorten the sequence seen by the recurrent stage, and test whether this preserves detection performance.

The six position channels are split into the three horizontal--vertical pairs of each BPM, and each pair is passed through an independent stack of three Conv1D layers with ReLU activations~\cite{nair2010relu} (Fig.~\ref{fig:cnn_architecture}, parameters in Table~\ref{tab:params}). In the rolling-normalized variant, each channel is first divided by its trailing baseline using the ratio of Eq.~(\ref{eq:rolling_norm}). The cascaded strides reduce the temporal dimension by a factor of 12, from 3000 samples to 250 covering the same time period, while the filters build an increasingly abstract representation that still captures signs of beam instability. The remaining LSTM layer is otherwise unchanged, aside from this shorter input window.

The convolutional front end increases the number of trainable parameters from $\sim$53,000 to $\sim$137,000, but reduces the model's per-window inference time by a factor of $\sim$13, both from the shorter recurrent pass over 250 rather than 3000 steps and from the parallelizable convolutional stacks over the three sensor pairs. Even with this reduction, meeting the $\sim$200\,MHz budget on FPGA/RFSoC fabric is not guaranteed by this speedup alone. Establishing whether either model meets that budget in absolute terms requires hardware-specific profiling planned in future work and discussed in Section~\ref{sec:compute}.

\section{Parameter optimization}\label{sec:optimization}

The detection algorithms of Section~\ref{sec:methods} have tunable parameters at two different levels. Model parameters, meaning window lengths, noise covariances, network architecture, and training settings, go into the anomaly score itself, so changing them requires re-training or re-running a model over the samples. Operating-point parameters, including the detection threshold, the pooling function across channels, and how many consecutive samples are needed to trigger, only act on the computed anomaly-score streams and can be swept without re-training.

Both levels are tuned against the figure of merits defined below, and every model sees the same set of events. DeepLSTM and the rolling-normalized hybrid Conv1D-LSTM were each retrained from 5 independent seeds on the same calibration pool to assess sensitivity to training stochasticity. Tables~\ref{tab:stage1} and \ref{tab:stage2} report the resulting range, and Fig.~\ref{fig:stage1_scatter} plots the average across seeds.

\subsection{Figure of merit and false alarms}\label{sec:fom}

The purpose of the oscillation-based models is to fire before the charge-based abort sensors. An operating point is therefore scored by the mean lead time $\Delta t$ (Section~\ref{sec:methods}) over all events not classified as false alarms by the criteria below.

A sufficiently low threshold will also fire on fluctuations that are indistinguishable from those seen in baseline, quiet-beam data: false alarms, and these must be accounted for in the figure of merit. For a trigger well before the reference abort, there are two possibilities: either the anomalous beam dynamics persist up to the abort, in which case we treat the trigger as a genuine early detection, or the beam relaxes back to ordinary behavior for an extended period before the current is eventually lost, in which case we treat it as unrelated and count it as a false alarm.

This distinction is not clean in practice: with limited quiet-beam data on tape and only approximately 100 turns of buffer before each abort, it is often difficult to say definitively whether a given early oscillation is related to the eventual loss. Nevertheless, we define false alarms concretely as follows. The relaxation level is set to the 95th percentile of the pre-trigger score, and a trigger is classified as a false alarm if the score drops below this level and stays low for at least two turns.\footnote{Brief spikes of up to three consecutive above-level samples are tolerated and do not reset the relaxation period.} Triggers that remain elevated through to the reference detection are counted as genuine early detections and enter the figure of merit with their full lead time. Others are excluded from the lead-time average and instead increment a per-model false alarm count $n_\text{FA}$.

\subsection{Model parameters}\label{sec:model_params}

Table~\ref{tab:params} summarises the tunable parameters of each model and which remain free after evaluation. Window lengths and smoothing windows for the classical detectors were set by coarse grid scans targeting the figure of merit at a fixed false alarm budget. The Kalman noise covariances were determined by grid search on stable beam data, and the LSTM architecture and training hyperparameters by grid search on forecasting error, i.e.\ without reference to detection performance. The cap on training sequences was set by compute budget. All parameters are frozen before the threshold optimization described next.

\begin{table*}[t]
\caption{Fixed parameters and trigger settings of the detection models.
Fixed parameters are chosen before evaluation, while the threshold and pooling
function are selected from the stored anomaly scores as described in
Section~\ref{sec:threshold_opt}. Windows are given in samples unless stated otherwise.}
\label{tab:params}
\centering
\small

\begin{tabular}{@{}l p{8.7cm} l@{}}
\toprule
Model
& Optimized on training set
& Optimized separately \\
\midrule

Sliding-window variance
& Variance window 500; baseline from first $5\times10^4$ pre-onset samples;
smoothing 500; persistence 1
& $k$, pooling \\

\quad rolling-normalized
& As above, applied to rolling-normalized channels (Eq.~\eqref{eq:rolling_norm}); baseline-ratio window 10 turns;
smoothing 500; persistence 10
& $k$, pooling \\

\addlinespace

Kalman velocity
& $Q=10^{-5}$; $R=10^{-4}$; velocity-mean window 10;
smoothing 500; persistence 50
& $k$, pooling \\

\addlinespace

DeepLSTM
& Sequence length 3000; two LSTM layers with 64 units each; dense layer 32;
dropout 0.2; learning rate $10^{-4}$; batch size 256; early stopping with
patience 5; at most $5\times10^5$ training sequences from the 35-event
calibration pool; residual clip 30; smoothing 500; persistence 1
& $k$, pooling \\

\addlinespace

Hybrid Conv1D--LSTM
& Conv1D filters $(32,64,64)$, kernels $(7,5,3)$, and strides $(2,3,2)$; evaluation stride 50; shared LSTM with 64 units; dense layer 32; dropout 0.2; training sequences from the 35-event calibration pool; persistence 1
& $k$, pooling \\

\addlinespace

\quad rolling-normalized
& As above, input rolling-normalized over 10 turns (Eq.~\eqref{eq:rolling_norm})
& $k$, pooling \\

\botrule
\end{tabular}
\end{table*}

\subsection{Threshold and pooling optimization}\label{sec:threshold_opt}

Once a model's anomaly-score streams are stored for every event, two degrees of freedom remain to be optimized post hoc: the detection threshold $\tau = \mu_\text{cal} + k\,\sigma_\text{cal}$ from Section~\ref{sec:methods}, and the channel-pooling function. The trigger persistence, \textit{i.e.} the minimum consecutive above-threshold run required to declare a trigger, is instead fixed per model ahead of evaluation, at the model-specific values stated in Section~\ref{sec:methods} and collected in Table~\ref{tab:params}, so a single noise spike can never fire a trigger without entering the sweep itself. The calibration statistics $\mu_\text{cal}$ and $\sigma_\text{cal}$ are computed per model from the stable-beam portions of a fixed pool of 35 calibration events, from which the DeepLSTM and hybrid models also draw their training sequences. Events that show significant oscillation during the baseline portion of the recording are excluded from the training sample. For every multichannel model, $k$ is swept from 2 to 100 jointly with the pooling function: maximum, mean, or root-mean-square across channels\footnote{Here $k$ is the number of calibration standard deviations used to define the threshold, but it should not be interpreted as a Gaussian significance. The pooled scores are strongly right-skewed, and the large number of samples considered in the sweep can require large values of $k$ to suppress false alarms.}.

\section{Results}\label{sec:results}

We present results for the LER, where the large majority of SBL events occurred in 2024 (Section~\ref{sec:sample}).
The 38 HER events are not evaluated separately because the sample is too small for reliable training and threshold calibration, and the LER models do not transfer directly between rings.
Of the 208 events used in the analysis (Section~\ref{sec:sample}), we use a calibration sample of 35 events to compute the statistics $\mu_\text{cal}$ and $\sigma_\text{cal}$ introduced in Section~\ref{sec:threshold_opt}. These events are held out from the validation sample. Alongside the mean and median time advantage over the BCM reference trigger, we quote two rates: an event counts as \textit{early} if the model triggers before the BCM reference and the trigger is not classified as a false alarm, and as \textit{safe} if it triggers at least $28\,\mu\mathrm{s}$ ahead. The second cutoff approximates the worst-case abort latency, so that a safe detection leaves enough time to dump the beam before the lost current reaches the IP.

The quiet-beam noise level varies substantially between events. Across the 35 calibration events, the mean pre-trigger DeepLSTM score ranges from 0.95 to 3.16. A threshold derived from the pooled calibration sample may depend on the particular events included, rather than representing a universal quiet-beam noise level. We therefore report two operating points for every model. The first calibrates $k$ and the pooling function for each model on 35 calibration events, and is afterwards evaluated on the remaining 173 events, which never enter the selection. The second optimizes $k$ and the pooling function on all 208 events at fixed maximum $n_{FA}$ and reports performance on those same events. The first measures how well a threshold transfers to unseen aborts, the second how much lead time each model can deliver once its threshold is matched to the sample.

\subsection{Held-out evaluation (Stage 1)}\label{sec:stage1}
\begin{figure}[h]
  \centering
  \includegraphics[width=\linewidth]{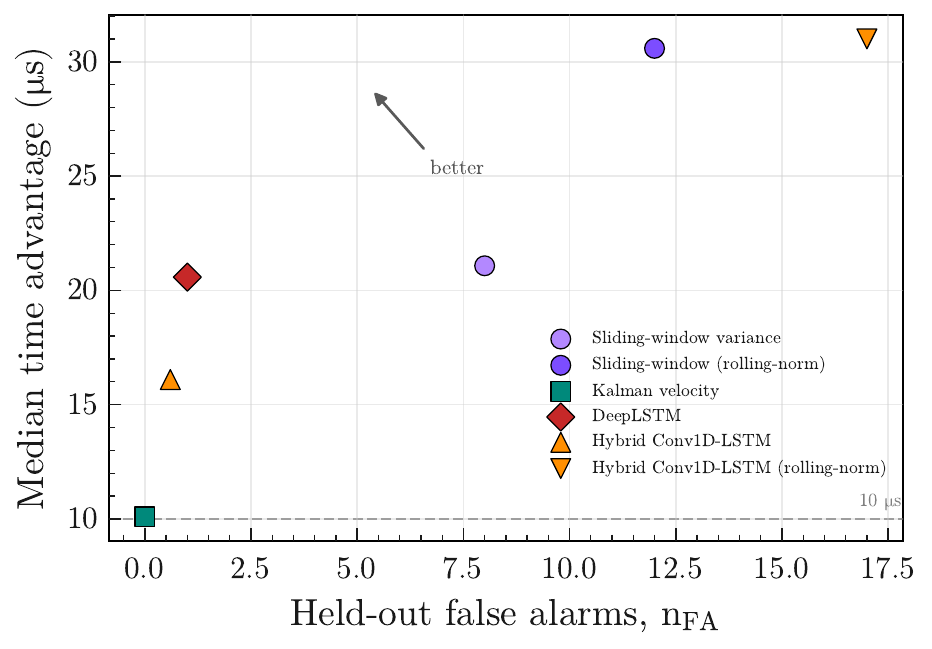}
  \caption{Median time advantage against false alarm count for each model, at the operating points of Stage 1 (Section~\ref{sec:stage1}). The arrow marks the preferred direction, combining early triggering with few spurious aborts. The horizontal line marks a median lead of $10\,\mu\mathrm{s}$. DeepLSTM and Hybrid Conv1D-LSTM points show the average over 5 independently trained seeds.}
  \label{fig:stage1_scatter}
\end{figure}

\begin{table*}[t]
\caption{Model comparison under held-out evaluation. The sensitivity parameter $k$ and the pooling function are selected on the 35 calibration events alone, the resulting threshold is inflated by 50\% as a margin against the small calibration sample, and performance is evaluated on the remaining 173 events. Time advantages are quoted relative to the BCM reference detection. DeepLSTM and Hybrid Conv1D-LSTM  rows show the min--max range over 5 independently trained seeds.}
\label{tab:stage1}
\centering
\small
\setlength{\tabcolsep}{3.5pt}
\begin{tabular}{@{}l r r r r r l@{}}
\toprule
Model & Early ($\Delta t > 0$) & Safe $\ge 28\,\mu\mathrm{s}$ & Median ($\mu\mathrm{s}$) & Mean ($\mu\mathrm{s}$) & $n_\text{FA}$ & Pooling, $k$ \\
\midrule
Sliding-window variance           & 148/173 (86\%) & 56/173 (32\%) & 21.1 & 27.0 & 8  & rms, $k = 8.2$ \\
\quad rolling-norm variant        & 158/173 (91\%) & 87/173 (50\%) & 30.6 & 44.1 & 12 & max, $k = 6.8$ \\
Kalman velocity                   & 129/173 (75\%) & 20/173 (12\%) & 10.1 & 11.5 & 0  & rms, $k = 4.5$ \\
DeepLSTM (multivariate)           & 149--156/173 (86--90\%) & 50--56/173 (29--32\%) & 19.6--21.7 & 25.4--29.8 & 1  & max, $k = 3.0$ \\
Hybrid Conv1D-LSTM                & 127--136/173 (73--79\%) & 36--40/173 (21--23\%) & 15.5--16.6 & 16.8--20.4 & 0--1  & max, $k = 6.0$ \\
\quad rolling-norm variant        & 149--157/173 (86--91\%) & 83--90/173 (48--52\%) & 30.2--32.6 & 50.3--54.5 & 14--22 & mean, $k = 3.0$ \\
\botrule
\end{tabular}
\end{table*}
For each model, $k$ and the pooling function are chosen using the procedure of Section~\ref{sec:threshold_opt} on the 35 calibration events, selecting the operating point that maximises the mean time advantage. To be conservative against the small calibration sample being unrepresentative, the threshold is increased by 50\% before being applied to the 173 held-out events. Table~\ref{tab:stage1} lists the full numbers and Fig.~\ref{fig:stage1_scatter} places the models against one another.

The classical detectors already deliver substantial lead time over the charge-based reference. The sliding-window variance detector triggers early in 86\% of held-out events with a mean time advantage of $27.0\,\mu\mathrm{s}$. However, it reaches that performance at $n_\text{FA} = 8$, meaning that it would have dumped a healthy beam up to eight times across the 173 held-out events. Each such abort costs a refill from zero and the luminosity accumulated with it, and at this rate the cost of spurious aborts likely outweighs the earlier signals. Rolling normalization of the input channels increases the early rate further to 91\%, at a mean time advantage of $44.1\,\mu\mathrm{s}$, and $n_\text{FA} = 12$.

The Kalman velocity detector instead settles at a conservative operating point. It registers no false alarms on held-out data, but much of the useful lead time is removed along with them, leaving a mean time advantage of $11.5\,\mu\mathrm{s}$ and a safe rate of 12\%. Together, the classical detectors illustrate the trade-off between early detection and false-alarm suppression. Lower thresholds retain sensitivity to the precursor but increase the false-alarm rate, while higher thresholds reduce false alarms at the cost of lead time.

The hybrid Conv1D-LSTM reaches 73--79\% early using raw (non-normalized) channels and a mean time advantage of $16.8$--$20.4\,\mu\mathrm{s}$, at $n_\text{FA} = 0$--1. Rolling normalization has a bigger effect than for the sliding-window detector, lifting the mean time advantage to $50.3$--$54.5\,\mu\mathrm{s}$ and the safe rate from 21--23\% to 48--52\%, both the best of any model in this comparison. The false alarm count rises with it, from 0--1 to 14--22, placing the model furthest to the right in Fig.~\ref{fig:stage1_scatter}. This shows that the transformation exposes precursor structure beyond the raw channels, but the 35-event calibration sample is not sufficient to set a threshold that separates that structure from ordinary beam motion.

DeepLSTM is the model located furthest toward the upper left in Fig.~\ref{fig:stage1_scatter}. It reaches an early rate of 86--90\%, with a mean time advantage of $25.4$--$29.8\,\mu\mathrm{s}$ and a median of $19.6$--$21.7\,\mu\mathrm{s}$, on a single false alarm across the 173 events ($n_\text{FA}=1$ for every seed). Only the two rolling-normalized detectors exceed it on lead time, and both do so at far higher false-alarm counts, 12 and 14--22 against DeepLSTM's single false alarm. In Section~\ref{sec:far} we examine the single false alarm that remains.

\subsection{False alarm analysis}\label{sec:far}

The false-alarm label does not represent ground truth, but identifies triggers for which the anomaly score returns to its pre-trigger level before the reference detection. This can misclassify weak or slowly developing precursors, as illustrated by the single false alarm assigned to DeepLSTM.

\begin{figure*}[t]
  \centering
  \includegraphics[width=\linewidth]{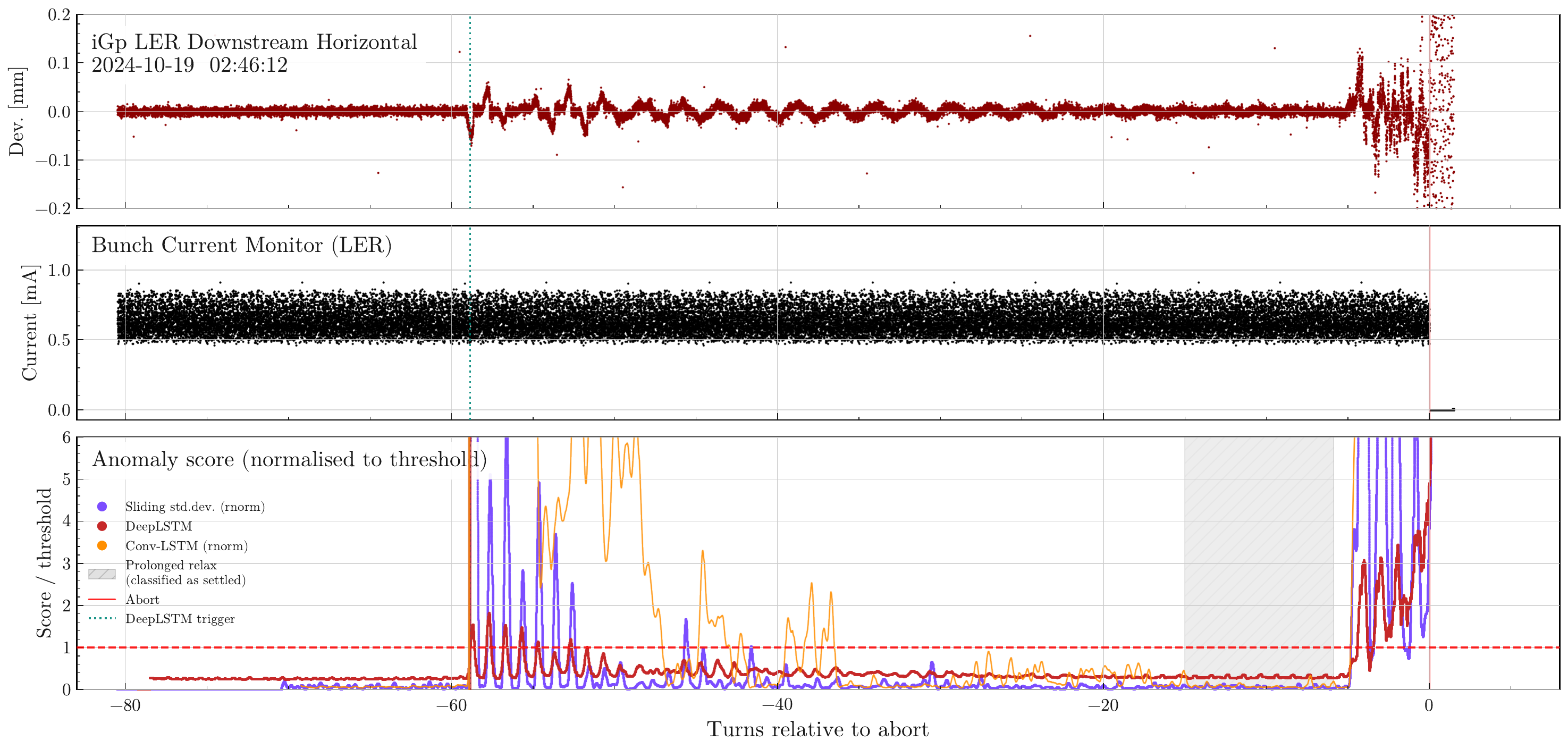}
  \caption{Sensor deviation, BCM bunch current, and anomaly score normalized to the detection threshold for the event of 19 October 2024, the single trigger registered as a false alarm by DeepLSTM. The shaded region marks the quiet run on the basis of which the trigger is classified as settled.}
  \label{fig:case_study}
\end{figure*}

The event, recorded on 19 October 2024 at 02:46:12, is shown in Fig.~\ref{fig:case_study}. Three detectors (the rolling-normalized sliding-window detector, DeepLSTM, and the hybrid Conv1D--LSTM) trigger on a sustained jitter approximately 59 turns before the abort. The score remains elevated for most of the remaining buffer, dipping below the settle level for about 9 turns immediately before the loss develops. This brief dip produces the false-alarm label, even though the beam does not fully return to its earlier quiet state. The event is therefore labeled as a false alarm, but the value $n_\text{FA}=1$ reported for DeepLSTM in Table~\ref{tab:stage1} should be interpreted as an upper bound on the number of genuinely spurious triggers.

\begin{figure}[h]
  \centering
  \includegraphics[width=\linewidth]{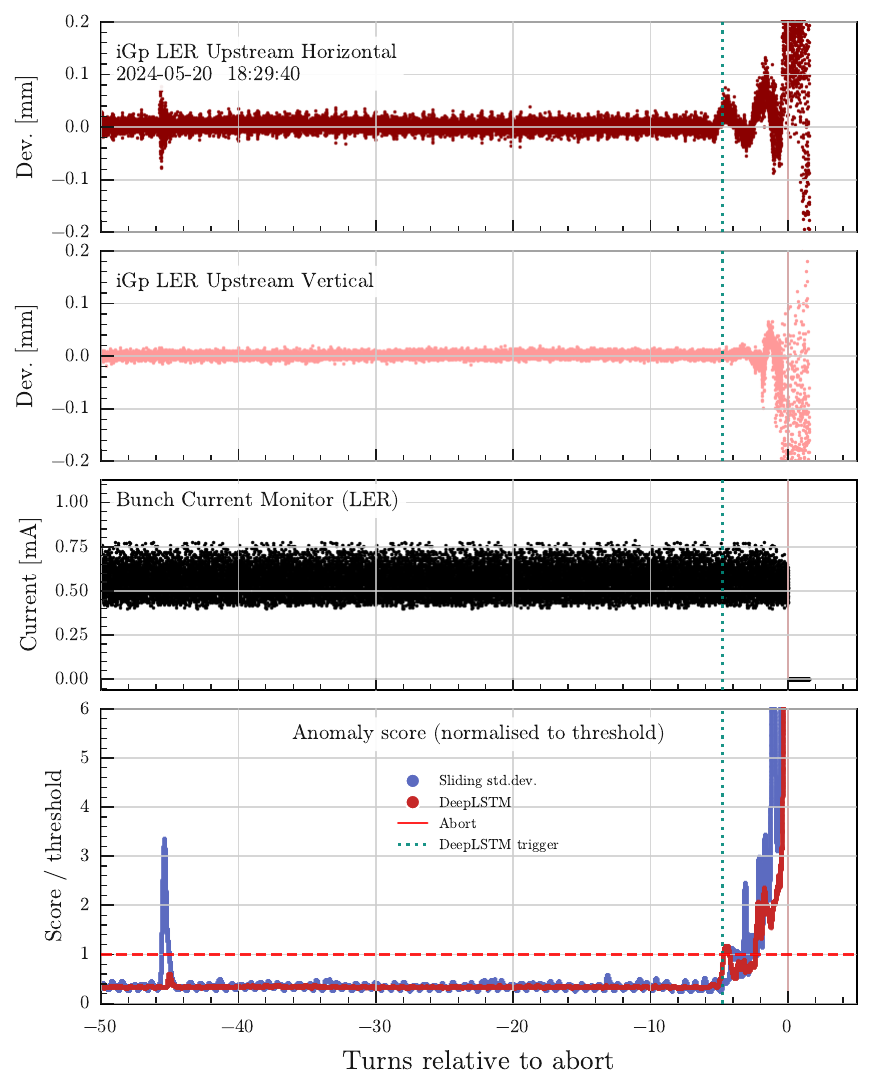}
  \caption{The SBL event on 20 May 2024, 18:29:40 registers as a false alarm for the Sliding window detector, but remains below threshold for the DeepLSTM model.}
  \label{fig:genuine_fa_example}
\end{figure}

Figure~\ref{fig:genuine_fa_example} shows a false alarm on the event of 20 May 2024 at 18:29:40. A brief spike in the upstream horizontal channel, with no corresponding trend in the vertical channel, triggers the sliding-window variance detector around 46 turns before the abort. After the spike, the bunch position returns to normal and all scores stay below threshold. The DeepLSTM model only crosses its threshold once the real precursor develops. Unlike the ambiguous case, the first trigger is clearly separated from the eventual precursor by a long quiet stretch.

\subsection{Fine-tuned evaluation on the full sample (Stage 2)}\label{sec:stage2}

Section~\ref{sec:stage1} withholds 173 of the 208 events from the search over $k$ so that the quoted performance is not fitted to the events on which it is measured. We now relax that constraint. While all major model parameters are retained from the training sample, we now select the sensitivity $k$ and the pooling function by optimizing on all 208 events. Table~\ref{tab:stage2} reports the result.

With the full sample available to the selection a false alarm budget becomes meaningful, and we allow up to one false alarm per model so that every model is quoted at the same tolerance and the lead times can be compared directly. Requiring $n_\text{FA} = 0$ leads to unnecessarily conservative working points, as Section~\ref{sec:far} shows that the classification is not definite and that a labeled trigger is not necessarily spurious. The thresholds that achieve $n_\text{FA} \le 1$ over all events lie above those of Section~\ref{sec:stage1} for the rolling-normalized models, with the sliding-window variant moving from $k = 6.8$ to $k = 67.5$ and the hybrid from $k = 3.0$ to $k = 69.5$.

The two multivariate models are now closely matched: the rolling-normalized hybrid Conv1D-LSTM registers 91--93\% of aborts early at a mean time advantage of $27.1$--$29.4\,\mu\mathrm{s}$ across seeds, while DeepLSTM registers 89--90\% at $28.5$--$29.3\,\mu\mathrm{s}$. The performance of the DeepLSTM on the hold-out and fine-tuned evaluation samples is fully statistically compatible, and it shows the best transfer between the two. With the full sample, the Kalman velocity detector reaches a competitive early-detection rate of 89\%, trailing only the two multivariate models. However, its mean lead time of $18.7\,\mu\mathrm{s}$ and safe rate of 23\% remain below those of the multivariate models. Its low computational cost therefore makes it a reliable fallback when consistency is more important than maximum lead time.
The plain sliding-window detector remains the weakest of the set at 62\% early, limited by how tightly its false-alarm count can be controlled. Its rolling-normalized variant fares better at 79\% early, at $k=67.5$.

\begin{table*}[t]
\caption{Model comparison under fine-tuned evaluation. The sensitivity parameter $k$ and the pooling function are selected on all 208 events subject to $n_\text{FA} \le 1$, and performance is reported on the same events, with $\mu_\text{cal}$ and $\sigma_\text{cal}$ still taken from the 35-event calibration pool. No held-out split is applied. DeepLSTM and Hybrid Conv1D-LSTM rows show the min--max range over 5 independently trained seeds.}
\label{tab:stage2}
\centering
\footnotesize
\setlength{\tabcolsep}{3.5pt}
\begin{tabular}{@{}l r r r r r l@{}}
\toprule
Model & Early ($\Delta t > 0$) & Safe $\ge 28\,\mu\mathrm{s}$ & Median ($\mu\mathrm{s}$) & Mean ($\mu\mathrm{s}$) & $n_\text{FA}$ & Pooling, $k$ \\
\midrule
Sliding-window variance           & 128/208 (62\%) & 26/208 (12\%) & 12.7 & 11.3 & 1 & mean, $k = 43.0$ \\
\quad rolling-norm variant        & 164/208 (79\%) & 64/208 (31\%) & 21.0 & 20.8 & 1 & max, $k = 67.5$ \\
Kalman velocity                   & 186/208 (89\%) & 47/208 (23\%) & 15.0 & 18.7 & 1 & max, $k = 12.0$ \\
DeepLSTM (multivariate)           & 186--188/208 (89--90\%) & 68--71/208 (33--34\%) & 21.7--22.6 & 28.5--29.3 & 1 & max, $k = 4.0$ \\
Hybrid Conv1D-LSTM                & 148--165/208 (71--79\%) & 40--48/208 (19--23\%) & 15.6--16.3 & 16.0--20.2 & 0--1 & mean, $k = 6.5$ \\
\quad rolling-norm variant        & 190--194/208 (91--93\%) & 66--75/208 (32--36\%) & 21.0--22.8 & 27.1--29.4 & 1 & max, $k = 69.5$ \\
\botrule
\end{tabular}
\end{table*}

The two evaluations address different questions. Section~\ref{sec:stage1} tests whether thresholds calibrated on a small sample transfer to unseen aborts. Under this criterion, DeepLSTM provides the best balance between early detection and false alarms. Section~\ref{sec:stage2} instead estimates the best performance achievable when the threshold is selected on the full sample. In this case, DeepLSTM and the rolling-normalized hybrid Conv1D--LSTM perform similarly.

Because the Stage~2 thresholds are selected and evaluated on the same events, these results represent an upper bound rather than a validated operating point. The difference between the two stages therefore mainly reflects threshold stability: models whose anomaly-score scale is poorly constrained by the calibration sample require the largest changes in $k$. Establishing a deployable operating point would require validation on additional independent data, as discussed in Section~\ref{sec:fa_rate}.

\subsection{Severity correlation}\label{sec:severity}

Of the 173 held-out events used in Section~\ref{sec:stage1}, 24 correspond to aborts that also triggered a QCS quench. Using the Stage-1 operating points and splitting by this shows that quench events are detected early more often than non-quench events, but with a shorter mean lead time (e.g. DeepLSTM: 100\% early at $16.5\,\mu\mathrm{s}$ mean for quench events, versus 88\% early at $32.0\,\mu\mathrm{s}$ mean for non-quench events). This is consistent with the most violent losses developing faster while leaving a comparatively unambiguous precursor, but the sample is small ($n=24$) and we do not treat this as conclusive.

We additionally test lead time against the diamond-sensor dose, the severity measure shown in Fig.~\ref{fig:dataset_currents}. Across all models, the Pearson correlation between dose and lead time is weak and not statistically significant ($|r| < 0.14$, $p > 0.05$ in every case). Deposited dose at the IP therefore does not detectably predict how early a given model triggers. This is plausibly because the sensor captures only local dose, while a more violent abort may deposit significant dose elsewhere in the ring that goes unmeasured.

\section{Discussion}\label{sec:caveats}

The results of Section~\ref{sec:results} are obtained offline from recorded buffers. Online use would require further validation of the false-alarm rate and computational cost, while additional sensors could improve detection performance.

\subsection{False-alarm rate}\label{sec:fa_rate}

The available data do not allow a true false-alarm rate to be measured. Each BOR buffer covers only approximately 100 turns around an abort, so false triggers during normal operation are not recorded. The reported quantity $n_\text{FA}$ therefore counts false alarms within these buffers rather than per unit of beam time and may underestimate the rate in operation.

This is particularly relevant because false-alarm performance separates DeepLSTM from the rolling-normalized hybrid in Section~\ref{sec:stage1}. A reliable comparison would require continuous quiet-beam data. This could be obtained by running the detectors in shadow mode, with triggers logged but not acted upon, or by recording additional buffers during normal operation. Such data could also help improve threshold calibration.

\subsection{Transfer to the 2025 -- 2026 runs}\label{sec:new-data}

All models in this study are optimized, trained and evaluated on 2024 data, before the vacuum-sealant removal described in Section~\ref{sec:sbl}. The BPMs have not been recalibrated since, and the quiet-beam noise on individual channels when using the previous calibration scheme has increased by up to a factor of three. We apply the detectors without retraining to the 77 SBL events of the 2025 and 2026 runs for which correctly processed BOR data were available at the time of this study. Only calibrations that require quiet-beam data are refit, using the first 20 events of the new runs, and performance is reported on the remaining events.

With the 2024 operating points applied unchanged, the detectors operating on raw amplitudes degrade substantially (Kalman: 35\% of events ahead of the reference), while the rolling-normalized detectors, which divide each channel by its own trailing baseline, perform roughly equivalently (hybrid Conv1D--LSTM: 75--84\% across seeds). Rescaling each channel so that its quiet-beam level matches that of 2024 restores the rolling-normalized sliding-window detector to 88\% of events ahead of the reference, compared with 91\% in 2024. Lead times are shorter than in 2024, with a median of 11--14\,$\mu$s, which may reflect faster loss development after the intervention but could also be a consequence of the increased noise level with incompatible calibrations.

These results indicate that the precursor persists after the vacuum sealant removal, and that the rolling-normalized detectors in particular can be kept operational across hardware and machine changes. New calibrations of the position monitors could improve the numbers for future run periods.
\subsection{Computational feasibility}\label{sec:compute}

The quoted lead times are useful only if the anomaly score can be evaluated fast enough online. The sliding-window and Kalman detectors require only simple operations per sample and are therefore the least computationally demanding. The hybrid Conv1D--LSTM reduces the sequence length by a factor of twelve before the recurrent layer, while DeepLSTM processes the full 3000-sample sequence and is expected to be more expensive.

A quantitative hardware study was not completed because the required vendor tools were unavailable. Resource use and latency on the ZCU111 and ZCU208 platforms therefore remain to be established. This is important because DeepLSTM and the rolling-normalized hybrid achieve similar performance in Section~\ref{sec:stage2}, but may differ substantially in deployment cost.

Quantization-aware training offers a route to the required throughput. Libraries such as QKeras~\cite{qkeras} and HGQ~\cite{hgq} for Keras, and Brevitas~\cite{brevitas} for PyTorch, train networks with low-precision fixed-point weights and activations, and hls4ml~\cite{hls4ml} converts them to FPGA firmware. Networks with ternary weights~\cite{bitnet} reduce the arithmetic further, and their precision limits for physics applications have been studied in Ref.~\cite{bithep}. Applying these methods to the hybrid Conv1D--LSTM, whose shortened recurrent stage is better suited to unrolled hardware implementations, is a natural next step.

\subsection{Additional sensors}\label{sec:sensors}

All six channels used in this study are measured in the Fuji section. Adding recorders at other positions around the ring could improve detection efficiency and lead time by capturing precursors that appear weakly or late at Fuji. Comparing signals across locations may also help identify where an event originates.

Some of the required instrumentation is already available: one of the RFSoC-based LER recorders installed in 2024 is located in the D05 section, and HER units are planned for 2027~\cite{nomaru2025}. These data were not included because the recorders had not yet been rigorously validated and calibrated for the 2024 runs.

\section{Summary}\label{sec:summary}

Sudden beam losses at SuperKEKB develop on a timescale comparable to the latency of the abort system. This study investigates whether turn-by-turn transverse position data from the Bunch Oscillation Recorders can provide earlier warning. Four detectors were evaluated on 208 LER events from the 2024 run using their lead time over a bunch-current-monitor reference and their false-alarm performance.

All four detectors anticipate the reference on most events, with typical lead times of several tens of microseconds. The comparison shows that the BOR data contain useful early-warning information, while the choice of detector determines the balance between lead time, false alarms, and computational cost.

The Kalman detector is lightweight and, when calibrated on the full sample, reaches a competitive early-detection rate of 89\%, although its lead times remain shorter than those of the forecasting models. DeepLSTM transfers most reliably from the 35-event calibration sample, reaching an early rate of 86--90\% with one labeled false alarm. The rolling-normalized hybrid Conv1D--LSTM achieves the longest lead times, but its threshold transfers less reliably and produces substantially more false alarms when calibrated on held-out events. When both models are tuned on the full sample, their performance is similar.

Measurements from recorders at additional locations around the ring may further improve detection efficiency and lead time. Before online deployment, however, two practical challenges remain. The abort-centred buffers do not provide enough quiet-beam exposure to measure a true false-alarm rate, and the computational feasibility of the models on the intended hardware has not yet been quantified. These points could be addressed through shadow-mode operation, additional quiet-beam recordings, and a dedicated hardware study.

\backmatter

\bmhead{Acknowledgements}
We thank the SuperKEKB operations and commissioning groups at KEK for providing the beam and for their feedback on this work, and the KEK instrumentation and feedback groups responsible for the Bunch Oscillation Recorders for making the data available. We thank the MBI IT group for providing computing resources. H.~Haigh and G.~Inguglia acknowledge financial support from the Austrian Science Fund (FWF) under the PAT program, grant DOI 10.55776/PAT1823724. P.~Horak is supported in part by the Natural Sciences and Engineering Research Council of Canada (NSERC).

\bibliography{cas-refs}

\end{document}